\documentclass[%
 aip,
 pof,
 amsmath,amssymb,
 preprint,
]{revtex4-1}

\usepackage{graphicx}
\usepackage{dcolumn}
\usepackage{bm}
\usepackage[utf8]{inputenc}
\usepackage[T1]{fontenc}
\usepackage{mathptmx}
\usepackage{amsmath,amssymb,amsthm}
\usepackage[english]{babel}
\usepackage{enumitem}
\usepackage[hidelinks]{hyperref}
\usepackage{booktabs}
\usepackage{array}
\usepackage[dvipsnames]{xcolor}

\graphicspath{{./}{../figures/main/}{../figures/controls/}}

\newtheorem{finding}{Empirical finding}
\newcommand{\R}{\mathbb{R}}
\newcommand{\E}{\mathbb{E}}

\newcommand{\diag}{\mathop{\mathrm{diag}}}
\newcommand{\dd}{\mathrm{d}}
\newcommand{\argmin}{\mathop{\mathrm{arg\,min}}}

\begin{document}

\title[Lagrangian ellipsoid diagnostics]{Lagrangian Ellipsoid Diagnostics in Rough Two-Dimensional Synthetic Flows:\\Aspect-Ratio Saturation and Reduced Modeling}

\author{Michael (Misha) Chertkov}
\email{chertkov@arizona.edu}
\affiliation{Graduate Interdisciplinary Program in Applied Mathematics, University of Arizona, Tucson, Arizona 85721, USA}
\affiliation{Department of Mathematics, University of Arizona, Tucson, Arizona 85721, USA}
\date{July 2026}

\begin{abstract}
We develop and test a Lagrangian methodology for extracting finite-cloud geometry and reduced dynamics from particle trajectories. A volume-filled cloud in a rough, incompressible, two-dimensional synthetic flow is represented simultaneously by a mass/gyration ellipse, which describes the particle-weighted bulk, and by a minimum-area enclosing ellipse, which describes the outer envelope and its boundary-supporting particles. The homogeneous Gaussian--H\"older velocity has finite Ornstein--Uhlenbeck temporal memory, and the moving cloud centroid is followed explicitly.

The principal empirical finding is unexpected: although the cloud is continually deformed by a non-smooth velocity field, the normalized aspect ratios of both ellipses reach broad, order-one distributions that are approximately stationary when conditioned on cloud scale. The mass ellipse is more elongated, whereas the enclosing ellipse covers a substantially larger radial envelope and has the more scale-stable conditional shape distribution.

To understand this saturation, we compare two finite-cloud descriptions of the velocity gradient. A gradient averaged spatially over the enclosing ellipse predicts more aligned stretching than the particles actually experience. The corresponding finite-cloud gradient correction provides most of the negative contribution that offsets aligned stretching; a smaller correction remains specific to representing the evolving outer cloud by an enclosing ellipse.

Finally, we demonstrate a data-to-model workflow. The measured time series is transformed to regular intrinsic variables, a constrained hierarchy of finite-lag stochastic models is fitted, and the models are compared on independent held-out realizations. The fitted coefficients and detailed stochastic equations are specific to this synthetic experiment. The intended transferable contribution is the methodology: paired cloud geometries, finite-cloud coarse graining, intrinsic variables, a transparent correction decomposition, and held-out validation for future direct numerical simulations and experiments.
\end{abstract}

\maketitle

\section{Introduction}
\label{sec:intro}

Lagrangian dispersion is often summarized by scalar quantities such as pair separation or moments of velocity increments. These quantities are fundamental, but they do not describe the joint shape of a multi-particle cloud or how that shape is related to the velocity field acting across the cloud. Earlier cluster and tetrad studies established that multi-particle shape, orientation, and alignment contain information unavailable to pair statistics \cite{chertkov_lagrangian_1999,pumir_geometry_2000}. In three-dimensional direct numerical simulations, Biferale \emph{et al.} found a statistically self-similar shape regime for four-particle tetrahedra, with strongly elongated and nearly planar configurations \cite{biferale_multiparticle_2005}. Multiparticle geometry has also been used to diagnose entrainment and deformation near turbulent/non-turbulent interfaces \cite{watanabe_multiparticle_2016}. More recently, principal-component and singular-value analyses of dense particle clouds were used to resolve anisotropic deformation in Rayleigh--B\'enard convection \cite{ettel_lagrangian_2026}. These studies motivate a diagnostic that can describe both the mass-carrying interior of a cloud and its outer extent.

The present paper develops such a diagnostic and then uses the resulting geometric time series to build a low-dimensional stochastic description. Finite-scale and particle-perceived velocity gradients provide a natural bridge between multiparticle geometry and local deformation \cite{johnson_multiscale_2024,yang_dynamics_2023,zhang_caustics_2025}. The guiding principle is straightforward: retain exact geometry and kinematics where they are available, model only the unresolved part, and accept additional model structure only when it improves predictions on independent data. This is the same broad physics-informed philosophy discussed in Ref.~\cite{chertkov_mixing_2024}. The required inputs are particle positions and velocities, or a resolved velocity field from which they can be evaluated, so the construction can later be applied to numerical simulations and particle-tracking experiments.

The experiment used here is deliberately synthetic. We prescribe a homogeneous, isotropic, incompressible Gaussian--H\"older velocity field with adjustable spatial roughness and finite temporal memory. Such random flows isolate Lagrangian mechanisms in a controlled setting, in the same methodological spirit as the Kraichnan model \cite{kraichnan_small-scale_1968,chertkov_normal_1995,gawedzki_anomalous_1995,shraiman_anomalous_1995}. The model does not contain Navier--Stokes dynamics, coherent vortices, or pressure-mediated interactions. It is therefore a first test of the methodology, not a quantitative model of a particular laboratory or geophysical flow.

The central empirical result is saturation of normalized cloud anisotropy in a non-smooth velocity field. A smooth affine flow provides only a simple reference: the mass ellipse and the minimum-area enclosing ellipse are transformed by the same deformation matrix and consequently have identical aspect-ratio dynamics, with exponential growth under sustained stretching \cite{batchelor_small-scale_1959,chertkov_statistics_1995,balkovsky_universal_1999,falkovich_particles_2001}. In a non-smooth flow, by contrast, particles at different locations experience different local deformations. The cloud is no longer an affine image of its initial shape, and the bulk and outer-envelope ellipses become inequivalent. It is then not evident that either normalized shape should approach a scale-independent statistical regime.

The simulations show that both do. The minimum-area enclosing ellipse develops a broad, order-one aspect-ratio distribution whose conditional mean and shape are approximately stationary over an extended range of cloud scales. The mass/gyration ellipse is more elongated but also shows no resolved continuing growth. At the same time, the enclosing ellipse reaches substantially farther from the center. Thus the two ellipses share the surprising saturation phenomenon while describing different geometric layers of the cloud.

The next question is how the positive stretching seen by the enclosing ellipse is balanced. We compare a velocity gradient averaged over the enclosing region with the best affine map (coarse-grained velocity gradient) fitted directly to the particle velocities. Their difference accounts for most of the negative correction to aligned stretching. A smaller residual remains after the particle-fitted map is used and is specific to the outer-envelope representation. This separation is useful because it distinguishes a mismatch in coarse graining from the additional effect of summarizing an irregular cloud by its enclosing ellipse.

The final part of the paper asks whether the measured geometric variables admit a useful reduced stochastic model. Recent work has used data-driven stochastic and dynamical models for Lagrangian velocity-gradient evolution \cite{das_data-driven_2024,carbone_tailor-designed_2024} and has emphasized the role of memory in reduced Lagrangian particle dynamics \cite{dewit_mori-zwanzig_2026}. Here we fit a hierarchy of finite-lag models in intrinsic variables and evaluate them on eight realizations excluded from fitting. The selected model is a dataset-specific statistical surrogate, not a universal closure. The methodological lesson is the sequence itself: define meaningful geometric objects, expose exact kinematics, estimate only unresolved terms, and use held-out tests to decide whether additional couplings or state variables are justified.

The paper follows this sequence. Section~\ref{sec:objects} defines the two cloud geometries, two finite-cloud velocity gradients, and the shape-production decomposition. Section~\ref{sec:flow} specifies the synthetic flow and numerical ensembles. Section~\ref{sec:saturation} establishes aspect-ratio saturation and the distinction between bulk and outer envelope. Section~\ref{sec:balance} analyzes the terms that balance aligned stretching. Section~\ref{sec:model} constructs and tests the reduced stochastic models. Section~\ref{sec:discussion} summarizes what is learned from the synthetic experiment and what should be tested next in realistic flows.

\section{Particle-cloud geometries and finite-cloud gradients}
\label{sec:objects}

\subsection{Moving cloud and mass/gyration ellipse}

Let $x_i(t)\in\R^2$, $i=1,\ldots,N$, be passive tracers. Their centroid and relative positions are
$X(t)=N^{-1}\sum_i x_i(t)$ and $y_i(t)=x_i(t)-X(t)$. The mass, or gyration, tensor is
\begin{equation}
G(t)=\frac{1}{N}\sum_{i=1}^N y_i(t)y_i(t)^\top.
\label{eq:gyration}
\end{equation}
We associate with it the Gaussian-equivalent mass ellipse
\begin{equation}
E_{\rm mass}(t)=\left\{y:y^\top(4G)^{-1}y\leq1\right\}.
\label{eq:mass-ellipse}
\end{equation}
The factor four makes this ellipse coincide with the boundary of a uniformly filled disk at initialization and does not affect its aspect ratio. Because every tracer contributes with equal weight to $G$, this ellipse describes the particle-carrying bulk. Equivalent principal-component and singular-value constructions are used in dense-cloud analyses of Rayleigh--B\'enard convection \cite{ettel_lagrangian_2026}.

Writing $\lambda_+(G)\geq\lambda_-(G)>0$, we define
\begin{equation}
r_{\rm mass}=(\det 4G)^{1/4},
\qquad
\sigma_{\rm mass}=\frac12\log\!\frac{\lambda_+(G)}{\lambda_-(G)}.
\label{eq:mass-observables}
\end{equation}
Thus $e^{\sigma_{\rm mass}}$ is the ratio of the principal semiaxes.

\subsection{Minimum-area enclosing ellipse}

The second object is the unique minimum-area ellipse containing all relative particle positions. We define its center $c_{\rm L}(t)$ and positive-definite shape matrix $g_{\rm L}(t)$ by
\begin{equation}
(c_{\rm L},g_{\rm L})
=\argmin_{c\in\R^2,\;g=g^\top\succ0}\log\det g
\quad\text{subject to}\quad
(y_i-c)^\top g^{-1}(y_i-c)\leq1,
\quad i=1,\ldots,N.
\label{eq:loewner-problem}
\end{equation}
The resulting ellipse is
\begin{equation}
E_{\rm L}(t)=\left\{y:(y-c_{\rm L})^\top g_{\rm L}^{-1}(y-c_{\rm L})\leq1\right\}.
\label{eq:loewner-ellipse}
\end{equation}
Since its area is $\pi\sqrt{\det g}$, Eq.~\eqref{eq:loewner-problem} is exactly the minimum-area enclosing problem. In convex geometry this object is called the \emph{L\"owner ellipse}; it is dual to the maximal-volume inscribed John ellipse \cite{john_extremum_1948,todd_minimum_volume_2016}. Below we use the descriptive terms ``minimum-area enclosing ellipse,'' ``enclosing ellipse,'' and the abbreviation MEE. The numerical solution uses the Khachiyan iteration \cite{khachiyan_rounding_1996}.

The L\"owner ellipse depends only on the convex hull of the cloud. A small set of boundary contact particles determines the optimum, so it probes the outer extent rather than the particle-weighted interior. Its scale and logarithmic aspect ratio are
\begin{equation}
r_{\rm L}=(\det g_{\rm L})^{1/4},
\qquad
\sigma_{\rm L}=\frac12\log\!\frac{\lambda_+(g_{\rm L})}{\lambda_-(g_{\rm L})}.
\label{eq:loewner-observables}
\end{equation}
For continuity with figure labels and data files, subscripts ``L'' and ``MEE'' are used interchangeably below.

\begin{figure*}[t]
\centering
\includegraphics[width=0.98\textwidth]{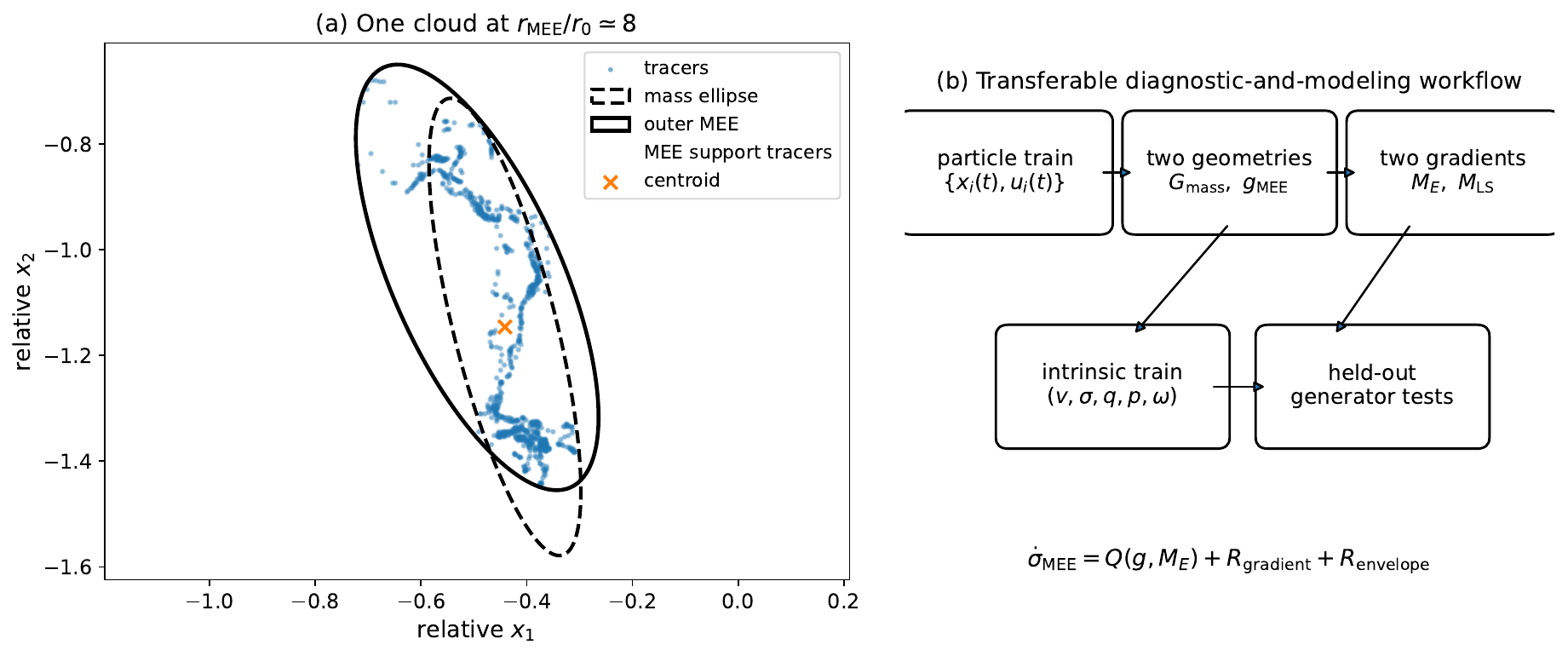}
\caption{\textbf{Two cloud geometries and the analysis workflow.} (a) A representative realization at $r_{\rm MEE}/r_0\simeq8$. The dashed mass ellipse follows the particle-weighted core; the solid minimum-area enclosing ellipse follows the cloud envelope. Open circles mark the three boundary particles with the largest normalized radii in this snapshot. (b) Particle trajectories are converted to two cloud geometries and two finite-cloud descriptions of the velocity gradient, and then to an intrinsic time series used for held-out model tests. The workflow, rather than the fitted numerical coefficients, is the component intended for transfer to future numerical and experimental datasets.}
\label{fig:methodology}
\end{figure*}

Fig.~\ref{fig:methodology} shows both ellipses for the same cloud. The mass ellipse follows an elongated core; the L\"owner ellipse follows boundary particles spread over a larger region. Comparing them separates deformation of the mass-carrying cloud from motion of its outer envelope.

\subsection{Two finite-cloud descriptions of the velocity gradient}

Once the two geometries are specified, we need a corresponding description of the velocity field acting on the cloud. The first is a spatial average of the instantaneous velocity gradient over the L\"owner ellipse,
\begin{equation}
M_{\rm L}(t)=\frac{1}{|E_{\rm L}(t)|}
\int_{E_{\rm L}(t)}\nabla u(x,t)\,\dd x.
\label{eq:M-L}
\end{equation}
This quantity answers the geometric question: what average linear deformation is present throughout the region enclosed by the cloud?

The second quantity is fitted to the velocities actually sampled by the particles. This construction is closely related to the perceived velocity gradient obtained from finite particle groups in laboratory and numerical turbulence studies \cite{yang_dynamics_2023,zhang_caustics_2025}. Define centered particle velocities
$\delta u_i=u(x_i,t)-N^{-1}\sum_j u(x_j,t)$ and the cross moment
$B=N^{-1}\sum_i\delta u_i y_i^\top$. The least-squares affine map is
\begin{equation}
M_{\rm LS}=B G^{-1}
=\argmin_M\sum_{i=1}^N\left|\delta u_i-My_i\right|^2.
\label{eq:M-LS}
\end{equation}
It answers a different question: which single linear map best represents the velocities of the finite set of particles? 

Direct differentiation of Eq.~\eqref{eq:gyration} gives the exact finite-cloud identity
\begin{equation}
\dot G=M_{\rm LS}G+GM_{\rm LS}^\top.
\label{eq:mass-exact}
\end{equation}
Thus the mass tensor $G$ has no unresolved kinematic term when it is paired with the particle-fitted map $M_{\rm LS}$. The difference of $M_{\rm LS}$ and $M_{\rm L}$ measures the mismatch between spatial averaging over the outer region and the affine deformation sampled by the particles. This distinction will be central to the shape balance in Sec.~\ref{sec:balance}.

\subsection{Intrinsic shape variables and a three-term balance}

For any positive shape matrix $g$, write
\begin{equation}
g={\cal R}(\theta_g)
\diag(e^{2v+\sigma},e^{2v-\sigma})
{\cal R}(\theta_g)^\top,
\label{eq:shape-decomposition}
\end{equation}
where ${\cal R}(\theta)$ is the two-dimensional rotation matrix. Then $e^v=(\det g)^{1/4}$ is the geometric-mean semiaxis, $e^\sigma$ is the aspect ratio, and $\theta_g$ is the direction of the major axis.

For an incompressible two-dimensional velocity gradient $M$, let $S=(M+M^\top)/2$ be its symmetric strain part. Write
\begin{equation}
S=A\,{\cal R}(\theta_S)\diag(1,-1){\cal R}(\theta_S)^\top,
\qquad A\geq0,
\label{eq:strain-decomposition}
\end{equation}
and define the relative angle and two signed strain components
\begin{equation}
\alpha=2(\theta_S-\theta_g),
\qquad
q(g,M)=2A\cos\alpha,
\qquad
p(g,M)=2A\sin\alpha.
\label{eq:qp}
\end{equation}
The component $q$ is the instantaneous production rate of logarithmic aspect ratio under an affine deformation: it is positive when the stretching direction is aligned with the major axis and negative when the stretching acts preferentially across it. The component $p$ measures the transverse part of the strain in the ellipse-aligned frame. We also use $\omega=(M_{21}-M_{12})/2$ for the local rotation rate. For $M_{\rm LS}$, which need not be trace free, $q$ and $p$ are computed from the deviatoric symmetric strain, $S^\circ_{\rm LS}=(M_{\rm LS}+M_{\rm LS}^\top)/2-[\operatorname{tr}(M_{\rm LS})/2]I$. The isotropic part affects only the area variable $v$ and cancels identically from the aspect-ratio rate $\dot\sigma$.

Combining Eqs.~\eqref{eq:mass-exact} and \eqref{eq:qp} gives
\begin{equation}
\dot\sigma_{\rm mass}=q(G,M_{\rm LS}).
\label{eq:mass-shape-production}
\end{equation}
No analogous identity closes the L\"owner-ellipse evolution using the spatial average $M_{\rm L}$. We therefore write the exact decomposition
\begin{align}
\dot\sigma_{\rm L}
&=q(g_{\rm L},M_{\rm L})+R_{\rm gradient}+R_{\rm envelope},
\label{eq:three-term}\\
R_{\rm gradient}
&=q(g_{\rm L},M_{\rm LS})-q(g_{\rm L},M_{\rm L}),
\label{eq:Rgrad}\\
R_{\rm envelope}
&=\dot\sigma_{\rm L}-q(g_{\rm L},M_{\rm LS}).
\label{eq:Renv}
\end{align}
The first correction, $R_{\rm gradient}$, measures the effect of replacing the spatially averaged gradient by the particle-fitted affine map. The second correction, $R_{\rm envelope}$, is what remains even after the best affine particle map is used; it includes the non-affine evolution of the cloud boundary and its projection onto a single enclosing ellipse. This terminology is deliberately descriptive: neither term is assumed in advance to be a universal relaxation law.

For reference, if $u(x,t)=A(t)x+b(t)$ is affine over the entire cloud, then $M_{\rm L}=M_{\rm LS}=A$, both $G$ and $g_{\rm L}$ transform by the same deformation matrix, and both corrections in Eqs.~\eqref{eq:Rgrad}--\eqref{eq:Renv} vanish. The two ellipses then have identical aspect-ratio dynamics and grow exponentially under sustained stretching. The remainder of the paper concerns the non-smooth case, where this affine equivalence is lost.

The definitions above are purely geometric and kinematic. We now specify the controlled random flow in which they are measured.

\section{Rough synthetic-flow experiment}
\label{sec:flow}

\subsection{Homogeneous incompressible Gaussian--H\"older field}

On a periodic square of side $L=2\pi$, we use the real full sine--cosine Fourier representation
\begin{equation}
u(x,t)=\sum_{k\in\Lambda_+}2\,\hat e_k^\perp
\left[a_k(t)\cos(k\cdot x)+b_k(t)\sin(k\cdot x)\right],
\label{eq:flow}
\end{equation}
where $\hat e_k^\perp$ is a unit vector perpendicular to $k$ and $\Lambda_+$ contains one member of each pair $\{k,-k\}$. Hence $\nabla\cdot u=0$ exactly. The real modal amplitudes are independent mean-reverting Gaussian Ornstein--Uhlenbeck processes. With stationary modal variance $V_k\propto|k|^{-(2\zeta+2)}$, they satisfy
\begin{align}
\dd a_k&=-\tau(k)^{-1}a_k\,\dd t+\sqrt{2V_k/\tau(k)}\,\dd W_{a,k},\\
\dd b_k&=-\tau(k)^{-1}b_k\,\dd t+\sqrt{2V_k/\tau(k)}\,\dd W_{b,k},
\label{eq:OU-modes}
\end{align}
with independent Wiener processes. The baseline correlation time is $\tau(k)\propto|k|^{-(1-\zeta)}$. Over the finite interval between the largest and smallest retained wavelengths, the second-order velocity increment scales approximately as $S_2(r)\sim r^{2\zeta}$.

The exponent $\zeta$ measures spatial roughness: typical velocity increments across a separation $r$ scale as $r^\zeta$. Values $\zeta<1$ correspond, in the ideal infinite-resolution limit, to a velocity field that is continuous but not spatially differentiable. The main ensemble uses $\zeta=1/3$, the dimensional exponent associated with Kolmogorov velocity-increment scaling. To determine whether the geometric findings depend strongly on this choice, we also simulate $\zeta=1/4$, $1/2$, and $2/3$. All four cases are therefore spatially non-smooth. The cross-roughness comparison appears in Sec.~\ref{sec:saturation}, Fig.~\ref{fig:bulk-envelope}(d), and Table~\ref{tab:ensembles}.

Eq.~\eqref{eq:flow} is spatially homogeneous. Particle positions are therefore integrated without wrapping, the moving centroid $X(t)$ is followed explicitly, and all cloud geometries are computed in relative coordinates. This full sine--cosine moving-cloud construction is used for every production and control ensemble.

\subsection{Production ensemble and exact ellipse average}

Let $\ell_{\rm uv}=2\pi/K_{\max}$ denote the smallest retained wavelength. The production ensemble uses $\zeta=1/3$, $K_{\max}=192$, $N=1000$, and $r_0=\ell_{\rm uv}$, with 24 independent flow realizations. The velocity is evaluated on a $768^2$ periodic grid and interpolated cubically to the particle positions.

At recorded times, $M_{\rm L}$ is evaluated without spatial quadrature. For an ellipse with center $c$ and shape matrix $g$,
\begin{equation}
\frac{1}{|E|}\int_E e^{ik\cdot x}\,\dd x
=e^{ik\cdot c}\,\Phi(\rho_k),
\qquad
\rho_k=\sqrt{k^\top gk},
\qquad
\Phi(\rho)=\frac{2J_1(\rho)}{\rho},
\label{eq:ellipse-fourier}
\end{equation}
where $J_1$ is the first-order Bessel function and $\Phi(0)=1$. Applying Eq.~\eqref{eq:ellipse-fourier} to Eq.~\eqref{eq:flow} gives
\begin{equation}
M_{\rm L}=2\sum_{k\in\Lambda_+}
\hat e_k^\perp k^\top
\left[-a_k\sin(k\cdot c_{\rm L})+b_k\cos(k\cdot c_{\rm L})\right]
\Phi\!\left(\sqrt{k^\top g_{\rm L}k}\right).
\label{eq:exact-averaged-gradient}
\end{equation}
This Fourier--Bessel expression is the exact spatial average for the truncated spectral field.

The nominal integration cap of the production runs is $T_{\max}=20$ in the nondimensional units of Eqs.~\eqref{eq:flow}--\eqref{eq:OU-modes}. In practice every production realization reaches a geometric stopping criterion first: either the geometric-mean MEE radius reaches $L/8$ or its major semiaxis reaches $L/4$. The realized stopping times range from $8.1$ to $15.9$ (mean $12.4$), and the usable trajectories reach $r_{\rm MEE}/r_0\simeq23$--$25$. Thus increasing the nominal run time alone would not create an indefinitely longer uncontaminated range; the useful range is limited by the spectral dynamic range and eventually by the periodic box. The production set is complemented by the controls summarized in Table~\ref{tab:ensembles}: other non-smooth exponents, spectral cutoffs, three initial radii, three temporal-memory choices, and nested particle subsets. All uncertainty bars in the main empirical figures are standard errors across independent flow realizations, rather than errors obtained by treating successive times from one realization as independent.

\begin{table}[t]
\caption{Numerical ensembles. The $\zeta=1/3$, $K_{\max}=192$ set is the production ensemble; the remaining sets test sensitivity to individual choices.}
\label{tab:ensembles}
\centering
\begin{tabular}{l c c c}
\hline\hline
ensemble & values & realizations & purpose\\
\hline
production & $K_{\max}=192$, $N=1000$ & 24 & extended saturation\\
roughness & $\zeta=1/4,1/3,1/2,2/3$ & $12,12,10,10$ & non-smooth comparison\\
cutoff & $K_{\max}=64,128,192$ & $8,12,24$ & spectral range\\
initial radius & $r_0/\ell_{\rm uv}=1/2,1,2$ & $8,12,8$ & initial-cloud sensitivity\\
temporal memory & turnover, constant, short & $12,8,8$ & temporal persistence\\
particle number & $N=100,300,1000$ & 6 nested & enclosing-ellipse convergence\\
\hline\hline
\end{tabular}
\end{table}

\subsection{Numerical extraction and finite-lag rates}

The L\"owner ellipse defined by Eq.~\eqref{eq:loewner-problem} is computed from the particle convex hull using the Khachiyan iteration \cite{khachiyan_rounding_1996}, with relative tolerance $10^{-5}$. The mass tensor is computed directly from Eq.~\eqref{eq:gyration}. Particle trajectories use time step $\delta t=0.01$ and are recorded every $0.1$.

Rates are evaluated over a finite lag $\Delta=0.1$: $\dot\sigma$ is represented by $[\sigma(t+\Delta)-\sigma(t)]/\Delta$, while each source term is averaged by the trapezoidal rule over the same interval. The balance is repeated at $\Delta=0.2$ and $0.4$. At $K_{\max}=192$, the interpolation audit gives a relative root-mean-square velocity error of $3.3\times10^{-3}$. Tightening the enclosing-ellipse solve changes $\sigma$ by $1.1\times10^{-2}$ on average. These numerical changes are small compared with the differences discussed below. Complete interpolation, ellipse-fit, affine-benchmark, particle-number, and finite-lag checks are supplied in the reproducibility package.

With the numerical experiment defined, we first ask the most direct question: how do the two normalized cloud shapes change as the cloud grows?

\section{Aspect-ratio saturation and bulk--envelope separation}
\label{sec:saturation}

\subsection{The principal surprise}

\begin{figure*}[t]
\centering
\includegraphics[width=0.98\textwidth]{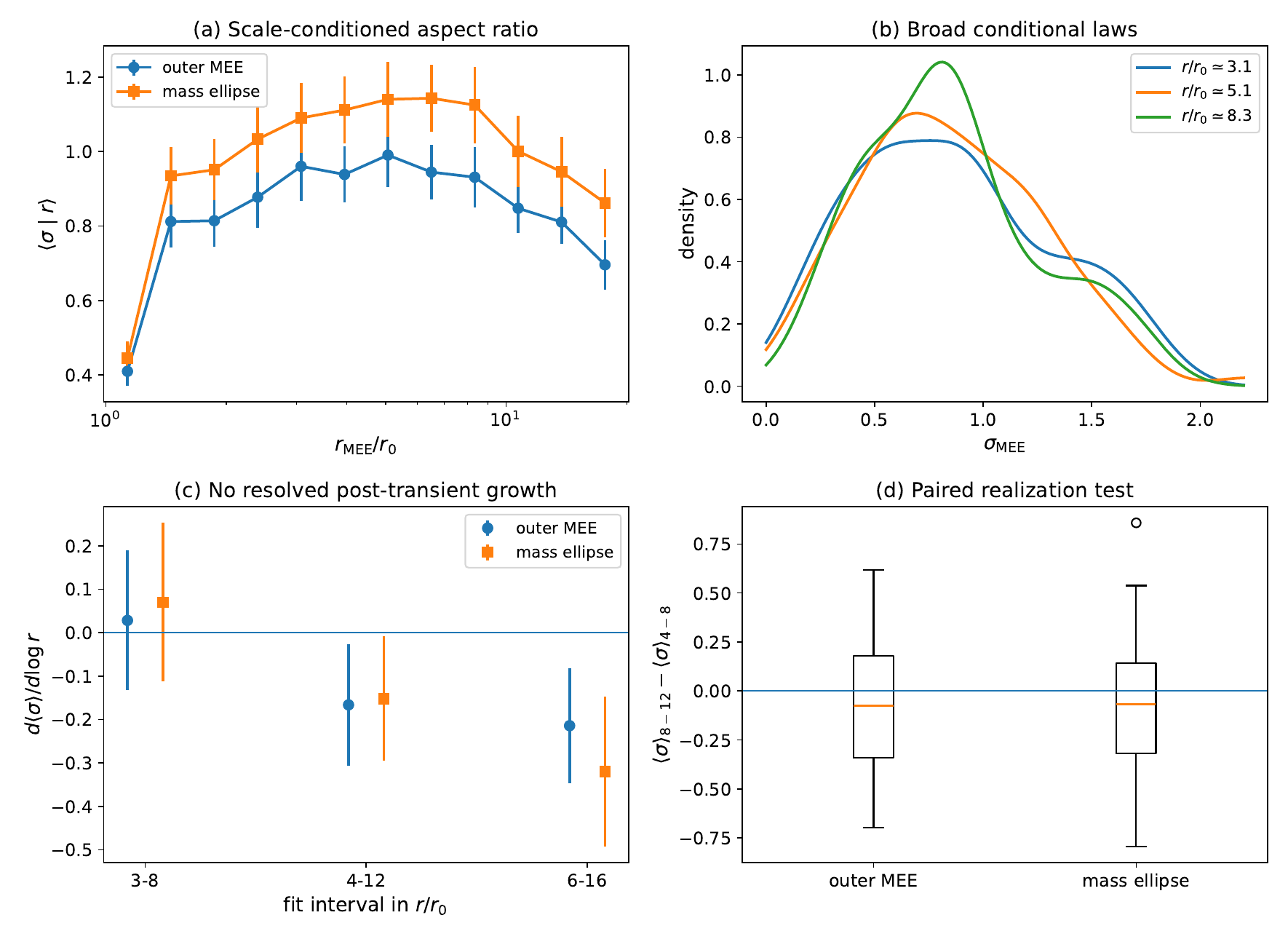}
\caption{\textbf{Aspect-ratio saturation in the non-smooth production ensemble.} (a) Conditional mean logarithmic aspect ratios of the outer enclosing ellipse and the mass ellipse. Error bars are standard errors across independent realizations. (b) Selected conditional distributions of the enclosing-ellipse aspect ratio remain broad but similar across separated scales. (c) Realization-level slopes over three post-transient intervals; none shows resolved positive growth. (d) Paired changes between $4\leq r/r_0<8$ and $8\leq r/r_0\leq12$.}
\label{fig:saturation}
\end{figure*}

Fig.~\ref{fig:saturation} presents the central result. After a short initial growth stage, the conditional means of both logarithmic aspect ratios remain of order one over a broad range of cloud scales. Here \emph{saturation} means approximate stationarity of ensemble shape statistics when conditioned on scale. It does not mean that an individual cloud approaches a fixed ellipse; individual realizations continue to fluctuate strongly.

Define the realization-level slope $\beta=\dd\langle\sigma\rangle/\dd\log(r_{\rm MEE}/r_0)$. On $3\leq r_{\rm MEE}/r_0\leq8$, $\beta_{\rm MEE}=0.028\pm0.161$ and $\beta_{\rm mass}=0.070\pm0.183$. On $4\leq r/r_0\leq12$, the slopes are $-0.166\pm0.140$ and $-0.152\pm0.144$, respectively; on $6\leq r/r_0\leq16$, they are $-0.214\pm0.133$ and $-0.320\pm0.172$. The paired late-minus-early changes are $-0.072\pm0.075$ for the enclosing ellipse and $-0.061\pm0.079$ for the mass ellipse. None of these tests supports continuing positive growth.

\begin{finding}[Scale-stationary rough-flow shapes]
In the tested non-smooth Gaussian--H\"older flow, both the outer-envelope and particle-weighted normalized shapes enter broad, order-one, approximately scale-stationary statistical regimes.
\end{finding}

This finding is not built into either ellipse construction and is established before any reduced model is fitted. It is the main surprise of the paper.

\subsection{Two distinct geometric layers}

Saturation does not make the two ellipses equivalent. On $3\leq r/r_0\leq8$, $\langle\sigma_{\rm MEE}\rangle=0.910\pm0.056$ and $\langle\sigma_{\rm mass}\rangle=1.070\pm0.068$, so the particle-weighted core is more elongated. Conversely,
$\langle\log(r_{\rm MEE}/r_{\rm mass})\rangle=0.338\pm0.018$, so the enclosing ellipse reaches substantially farther from the center. The mean anisotropy gap is
$\langle\sigma_{\rm mass}-\sigma_{\rm MEE}\rangle=0.160\pm0.041$.

\begin{figure*}[t]
\centering
\includegraphics[width=0.98\textwidth]{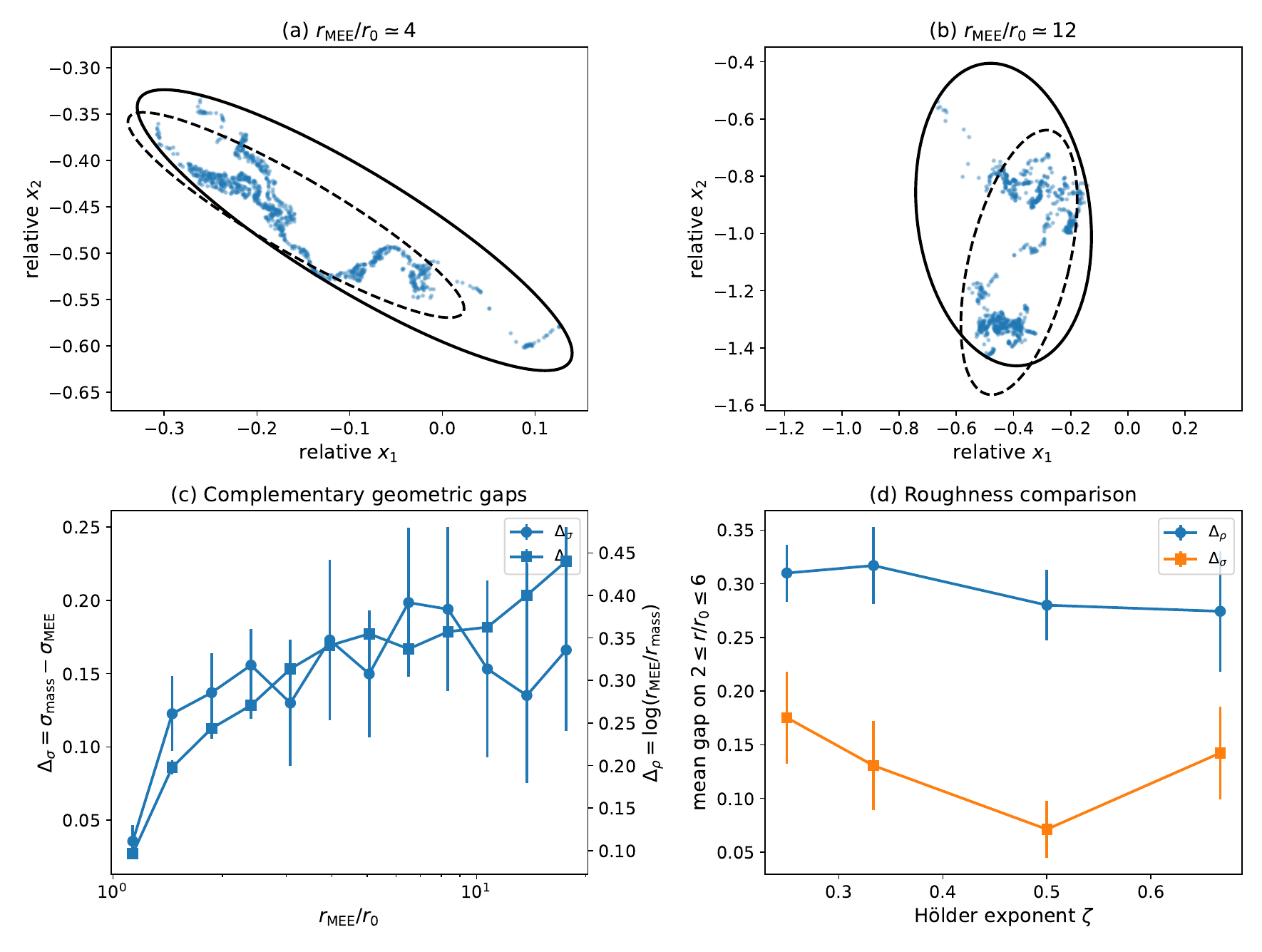}
\caption{\textbf{Bulk and outer-envelope diagnostics.} (a,b) Representative snapshots at two scales. The dashed mass ellipse follows the dense core, whereas the solid minimum-area enclosing ellipse follows the outer extent. (c) The anisotropy gap $\Delta_\sigma=\sigma_{\rm mass}-\sigma_{\rm MEE}$ and radial gap $\Delta_\rho=\log(r_{\rm MEE}/r_{\rm mass})$ remain positive over the post-transient interval. (d) Comparison across four non-smooth H\"older exponents. The radial gap is somewhat larger in the rougher cases, but the present uncertainty does not establish a monotone roughness law.}
\label{fig:bulk-envelope}
\end{figure*}

The enclosing-ellipse distribution also changes less across the central scale bins. A normalized pairwise 1-Wasserstein distance --- a measure of the separation between two one-dimensional distributions, computed pairwise among the three central logarithmic scale bins covering $3.5\leq r/r_0\leq7.4$ --- averages $0.127$ for the enclosing ellipse and $0.249$ for the mass ellipse. Fig.~\ref{fig:bulk-envelope} shows how the same distinction appears in individual clouds. Boundary particles are not discarded as statistical outliers: they are precisely the particles that define the outer geometric layer.

\subsection{Roughness and robustness}

On the matched interval $2\leq r/r_0\leq6$, the mean radial gaps $\Delta_\rho$ are $0.310$, $0.317$, $0.280$, and $0.274$ for $\zeta=1/4,1/3,1/2,2/3$. This ordering is consistent with a larger separation between leading particles and the bulk in rougher fields, but a realization-level bootstrap interval for the slope includes zero. We therefore report a robust bulk--envelope separation, not a universal monotone law in $\zeta$.

The enclosing-ellipse plateau survives every focused control. Changing $K_{\max}$, $r_0/\ell_{\rm uv}$, or temporal memory changes finite-range details and plateau levels but does not produce a statistically resolved positive post-transient slope. With nested $N=100,300,1000$ clouds, the mean enclosing-ellipse aspect ratios on $2\leq r/r_0\leq6$ are $0.959$, $0.903$, and $0.913$. Thus the normalized enclosing-ellipse shape is effectively converged by $N\simeq300$. The radial outer extent continues to increase with $N$, as expected for a statistic controlled by the most distant particles. Complete control plots are included in the reproducibility package.

The result established here applies to this controlled two-dimensional random-flow family. It is not asserted as a law for realistic three-dimensional flows. Earlier simulations of four-particle tetrahedra found a self-similar shape regime in homogeneous turbulence \cite{biferale_multiparticle_2005}, while recent convection simulations found strongly environment-dependent deformation histories for dense clouds \cite{ettel_lagrangian_2026}. These studies show that the required geometric measurements are meaningful, but they leave the present question open: do bulk and outer-envelope aspect-ratio distributions become scale stationary in realistic flows, and under what conditioning?

Having established saturation, we next examine the measured rates that maintain it.

\section{Aligned-strain production and finite-cloud corrections}
\label{sec:balance}

\begin{figure*}[t]
\centering
\includegraphics[width=0.98\textwidth]{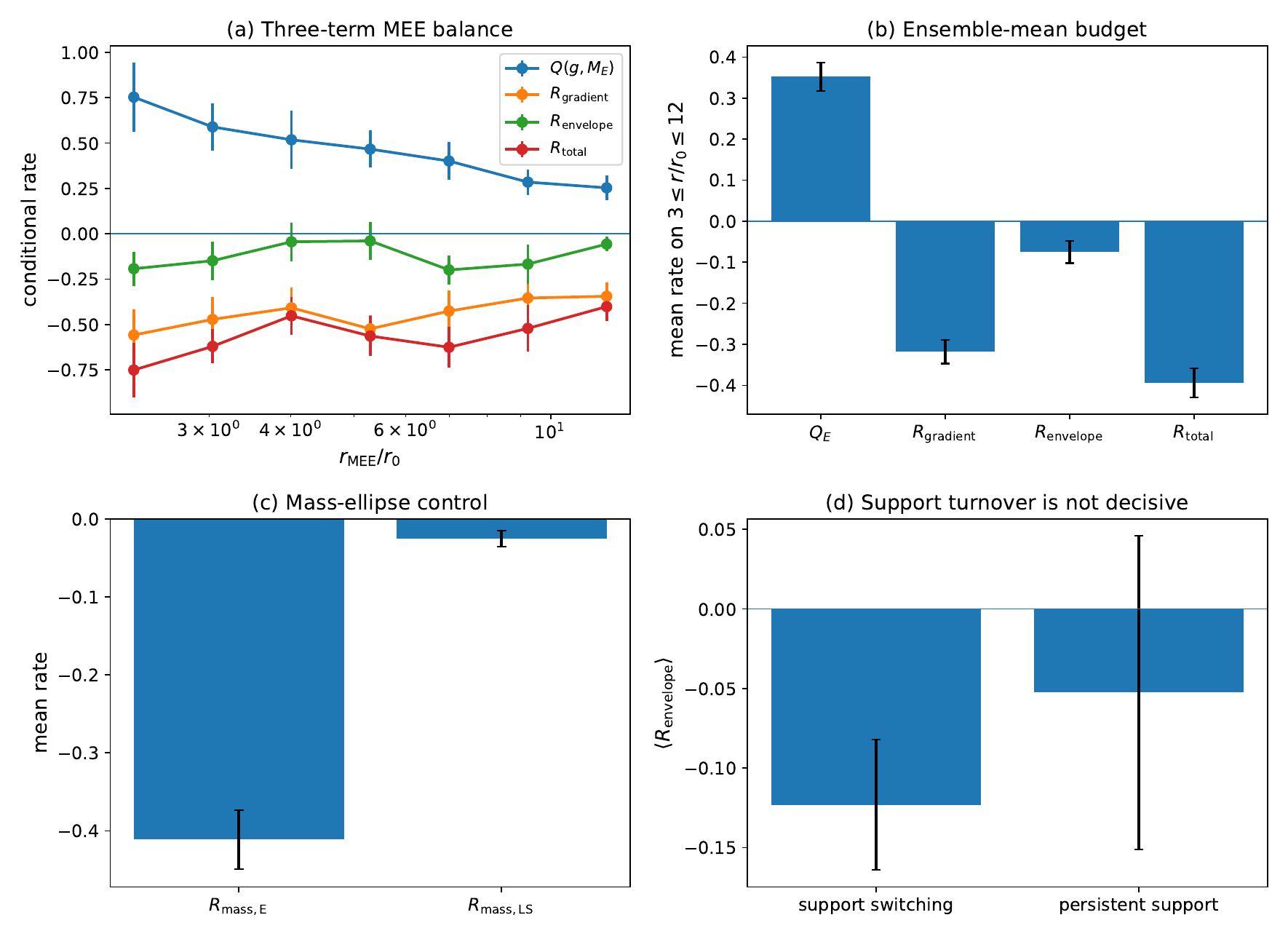}
\caption{\textbf{Shape-production/correction balance.} (a) Conditional rates in the production ensemble. (b) Ensemble means on $3\leq r/r_0\leq12$: positive aligned stretching measured over the enclosing region is offset mainly by the finite-cloud gradient correction and secondarily by the outer-envelope correction. (c) For the mass ellipse, the residual relative to $M_{\rm LS}$ is compatible with zero, as expected from Eq.~\eqref{eq:mass-exact}. (d) Intervals with changes in the tracked boundary particles and intervals with persistent tracked particles have statistically overlapping envelope corrections; turnover of these particles is therefore not established as the cause of saturation.}
\label{fig:balance}
\end{figure*}

Fig.~\ref{fig:balance} evaluates Eqs.~\eqref{eq:three-term}--\eqref{eq:Renv}. The first term is the aspect-ratio production predicted from the gradient averaged over the enclosing region. The next two terms show why that prediction does not equal the actual MEE rate.

At $\Delta=0.1$ and $3\leq r/r_0\leq12$,
\begin{align*}
\langle q(g_{\rm L},M_{\rm L})\rangle&=0.352\pm0.034,\\
\langle R_{\rm gradient}\rangle&=-0.318\pm0.029,\\
\langle R_{\rm envelope}\rangle&=-0.076\pm0.027.
\end{align*}
About $81\%$ of the mean negative correction is therefore explained by the difference between $M_{\rm L}$ and $M_{\rm LS}$: averaging the gradient over the entire enclosing region predicts more aligned stretching than the particle velocities support. The remaining $19\%$ is the outer-envelope correction. This smaller term is resolved from zero, and its sign and magnitude remain stable when the lag is changed from $0.1$ to $0.2$ and $0.4$.

The mass ellipse provides a direct control on this interpretation. Eq.~\eqref{eq:mass-exact} states that, at an instantaneous level, $M_{\rm LS}$ determines the mass-tensor evolution exactly. When the same finite-lag numerical procedure is applied to the mass ellipse, the residual relative to the spatial average $M_{\rm L}$ is negative, but the residual relative to $M_{\rm LS}$ is $\langle R_{\rm mass,LS}\rangle=-0.025\pm0.010$. This value is essentially independent of the lag ($-0.026$ at both $\Delta=0.2$ and $0.4$), identifying it as the empirical numerical floor of the finite-lag and trapezoidal evaluation at the stored recording cadence, and it is an order of magnitude smaller than $\langle R_{\rm mass,E}\rangle=-0.411\pm0.038$. Thus the large negative correction is not created by numerical differentiation alone, and the outer-envelope correction $\langle R_{\rm envelope}\rangle$ is approximately three times larger in magnitude than this empirical numerical floor. Replacing the regional average by the particle-fitted map reduces it to the numerical floor for the bulk ellipse, while a smaller correction remains for the outer ellipse because the MEE is not a material second-moment tensor.

The boundary particles are also dynamically distinct. The root-mean-square error of their velocities relative to the best affine cloud map is $1.526\pm0.030$ times the corresponding error for the remaining particles. We track the three boundary particles with the largest normalized MEE radii; their identities change frequently. However, on the same interval $3\leq r/r_0\leq12$, the mean envelope correction during intervals with a change in this tracked set, $-0.123\pm0.041$, is statistically indistinguishable from the value during intervals in which the set persists, $-0.053\pm0.099$. The boundary particles therefore contain information absent from the mass ellipse, but the present data do not identify turnover of their identities as the sole mechanism producing saturation.

\begin{finding}[Shape-production/correction balance]
Positive aspect-ratio production predicted from aligned strain over the enclosing region is offset mainly by the mismatch between that regional average and the affine deformation sampled by the particles, with a smaller additional correction associated with the evolving outer-envelope representation.
\end{finding}

This decomposition replaces the earlier practice of assigning every unresolved effect to one scalar ``relaxation'' term. It yields separate quantities that can be measured and compared in other synthetic flows, numerical simulations, and experiments.

The balance explains the average rate of change but not the temporal statistics of the measured variables. We therefore turn next to a reduced stochastic description.

\section{Physics-informed reduced modeling}
\label{sec:model}

\subsection{Objective, state, and model hierarchy}

The modeling objective is to construct a compact finite-lag transition model from the measured time series, retain the exact aligned-strain term in the shape equation, and determine on independent realizations which additional couplings are supported. The resulting coefficients and even the preferred architecture are properties of the present synthetic dataset; they are not proposed as universal equations.

We use the intrinsic state
\begin{equation}
y=(v,\sigma,q,p,\omega),
\qquad
v=\log(r_{\rm MEE}/r_0),
\label{eq:model-state}
\end{equation}
where $\sigma=\sigma_{\rm MEE}$ and $(q,p,\omega)$ are obtained from $g_{\rm L}$ and $M_{\rm L}$ through Eq.~\eqref{eq:qp}. Unlike the strain amplitude $A\geq0$ and an angle defined modulo $\pi$, the signed variables $q$ and $p$ have no positivity boundary or angle-wrapping discontinuity.

The scale coordinate is modeled by
\begin{equation}
\dd v=b_v(v)\,\dd t+\sqrt{Q_v(v)}\,\dd W_v.
\label{eq:vmodel}
\end{equation}
The shape equation keeps the measured aligned-strain source explicitly,
\begin{equation}
\dd\sigma=\left[q+a_0(v)+a_1(v)\sigma\right]\dd t
+\sqrt{Q_\sigma(v)}\,\dd W_\sigma.
\label{eq:sigmamodel}
\end{equation}
With $z=(q,p,\omega)^\top$, the remaining gradient variables obey, within each scale bin,
\begin{equation}
\dd z=\left[c(v)+L(v,\sigma)z\right]\dd t+B(v)\dd W_z,
\qquad Q_z(v)=B(v)B(v)^\top.
\label{eq:zmodel}
\end{equation}
Here $Q_v$ and $Q_\sigma$ are scalar noise intensities, and the symmetric matrix $Q_z$ gives the covariance per unit time of the random increments in $(q,p,\omega)$. The fitted functions are piecewise constant in five scale bins, which should be interpreted as a data-resolved approximation to smooth scale dependence rather than as five separate physical regimes.

The hierarchy tests progressively richer drift structures:
\begin{description}[leftmargin=2.1cm,style=nextline]
\item[M0] a null model with independent mean-reverting $q,p,\omega$ drivers and the systematic mean $q$ source suppressed;

\item[M1] independent mean-reverting intrinsic drivers with the mean $q$ source fitted from data;

\item[M2] a sparse coupled model in which $q$ depends on $(\sigma,q)$, $(p,\omega)$ form a coupled two-variable block, and the shape correction in Eq.~\eqref{eq:sigmamodel} is affine in $\sigma$;

\item[M3] a fully coupled linear drift in $(\sigma,q,p,\omega)$.
\end{description}
Counting drift and diffusion entries separately in each scale bin, M0, M1, M2, and M3 contain 15, 16, 20, and 26 fitted scalar values per bin, respectively.

\subsection{Finite-lag estimation and held-out design}

The 24 production realizations are divided before fitting: seeds 0--15 form the training set and seeds 16--23 form the held-out test set. Five scale bins are defined by quantiles of the training values of $v$. All bin boundaries, regression coefficients, and noise covariances are determined from the training realizations; the held-out realizations are used only to compare the completed models.

For a state $Y_t$, the effective drift and increment covariance at a finite lag $\Delta$ are
\begin{equation}
b_\Delta(y)=\frac{1}{\Delta}\E\!\left[Y_{t+\Delta}-Y_t\mid Y_t=y\right],
\qquad
Q_\Delta(y)=\frac{1}{\Delta}\operatorname{Cov}\!\left[Y_{t+\Delta}-Y_t\mid Y_t=y\right].
\label{eq:finite-lag-generator}
\end{equation}
These conditional moments define the Gaussian transition approximation used here. Because $\Delta$ is finite rather than infinitesimal, they should be viewed as effective finite-time coefficients; finite-sampling effects in such reconstructions are well known \cite{ragwitz_finite-time_2001}. We use $\Delta=0.1$, equal to the recording interval. For the shape equation, the aligned source $q$ is first averaged over the same interval and subtracted from $[\sigma(t+\Delta)-\sigma(t)]/\Delta$; the remainder is then regressed on $1$ and $\sigma$. For $z$, the three components of $[z(t+\Delta)-z(t)]/\Delta$ are regressed on the variables allowed by M0--M3.

A small ridge penalty stabilizes regressions when explanatory variables are correlated. The constant offset in each regression is left unpenalized, while the coefficients multiplying $\sigma$, $q$, $p$, or $\omega$ are penalized. The base multiplier is $10^{-3}$ for M0--M2 and $10^{-2}$ for the more highly parameterized M3, with normalization by the design-matrix trace inside each bin.

After fitting the mean increment, let $\epsilon_z$ and $\epsilon_\sigma$ denote the residual rates. The noise estimates are
\begin{equation}
Q_z=\Delta\,\operatorname{Cov}(\epsilon_z),
\qquad
Q_\sigma=\Delta\,\operatorname{Var}(\epsilon_\sigma).
\label{eq:diffusion-estimate}
\end{equation}
Sampling noise can produce a tiny negative eigenvalue in the estimated symmetric matrix. We therefore diagonalize $Q_z=V\diag(\lambda_j)V^\top$ and replace each $\lambda_j$ by $\max(\lambda_j,\varepsilon)$ with a small numerical floor $\varepsilon>0$. This guarantees that the covariance used to generate random increments is positive definite, as any physical covariance matrix must be.

The lag is checked using both predictive score and residual memory. For component $j$, define the standardized one-step prediction error
\begin{equation}
\eta_n^{(j)}=
\frac{Y_{n+1}^{(j)}-Y_n^{(j)}-\widehat b_j(Y_n)\Delta}
{\sqrt{\widehat Q_{jj}(Y_n)\Delta}}.
\label{eq:innovation}
\end{equation}
A well-specified one-step model should leave little correlation between consecutive $\eta_n^{(j)}$. At $\Delta=0.1$, the M2 lag-one correlations are $(0.125,0.035,-0.010,-0.058)$ for $(\sigma,q,p,\omega)$. At $\Delta=0.2$ and $0.4$, both these correlations and the predictive scores become worse. Sample counts, bin edges, fitted coefficients, and realization-bootstrap intervals are supplied in machine-readable tables and in the generator-identification notebook.

\subsection{Held-out model selection and interpretation}

\begin{figure*}[t]
\centering
\includegraphics[width=0.98\textwidth]{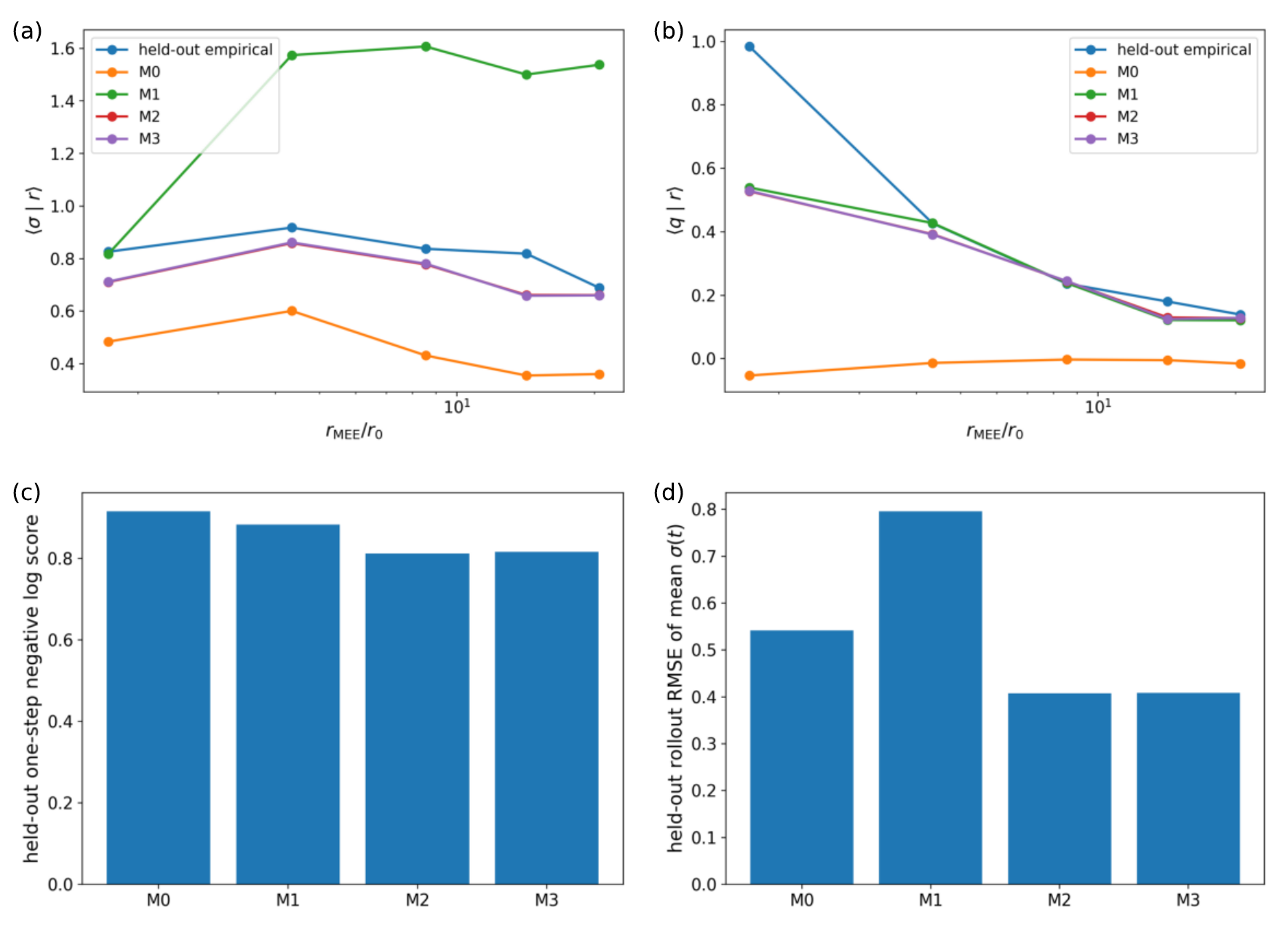}
\caption{\textbf{Held-out reduced-model comparison.} (a) Conditional mean aspect ratio and (b) aligned source on eight held-out realizations. (c) Held-out one-step negative log predictive score; smaller is better. (d) Held-out root-mean-square error of the mean $\sigma$ rollout when the empirical scale path is supplied. M2 is the simplest tested model that improves both short-time scoring and multistep shape statistics.}
\label{fig:model}
\end{figure*}

Fig.~\ref{fig:model} compares the hierarchy. The one-step score is the average Gaussian negative log predictive density,
\begin{equation}
\mathcal S=\frac12\left[\log\det(2\pi C)+e^\top C^{-1}e\right],
\label{eq:nll-score}
\end{equation}
where $e$ is the held-out prediction error and $C$ is the predicted increment covariance. A smaller value means that the model assigns higher probability to the observed next step while accounting for its predicted uncertainty.

The held-out one-step scores are $0.915$, $0.883$, $0.812$, and $0.816$ for M0--M3. The corresponding root-mean-square errors of the mean $\sigma$ rollout are $0.541$, $0.835$, $0.410$, and $0.411$. M2 is therefore selected: it improves both the null and independent-driver models, while the fully coupled M3 adds six fitted values per scale bin without measurable benefit.

An additional autonomous check fits Eq.~\eqref{eq:vmodel} from the training data and simulates the complete system in Eqs.~\eqref{eq:vmodel}--\eqref{eq:zmodel}. The held-out conditional-mean $\sigma(v)$ root-mean-square error is $0.095$ for M2, compared with $0.361$ for M0 and $0.747$ for M1. M3 gives $0.093$, an immaterial improvement relative to its six additional coefficients per scale bin and its slightly worse one-step score. This check confirms the M2 selection without supplying the empirical scale history.

We also tested explicit stochastic models of the tensor difference $M_{\rm LS}-M_{\rm L}$. The best such dynamic extension reduced the rollout error only from $0.410$ to $0.401$, while worsening the one-step score and the Chapman--Kolmogorov consistency test (from about $0.51$ to $0.79$). The tensor difference is therefore retained as an interpretive diagnostic but not added to the predictive state.

The selected generator remains approximate. Its normalized Chapman--Kolmogorov discrepancy is about $0.51$, and the standardized prediction errors retain weak temporal correlation. M2 is therefore a low-order statistical surrogate rather than a trajectory-accurate Markov model.

The comparison between M2 and M3 gives a useful modeling lesson. Adding every instantaneous linear coupling does not cure the remaining error; M3 can fit more coefficients but does not improve independent predictions. The limitation is therefore more likely to lie in the chosen instantaneous state or in the Gaussian, memoryless form of the noise than in the sparsity pattern of M2. Plausible next extensions include a short memory variable or delay coordinate, the measured non-affinity of the particle velocities, descriptors of the boundary-contact configuration, or state-dependent non-Gaussian increments. The explicit tensor-corrector test shows that simply appending $M_{\rm LS}-M_{\rm L}$ in its present form is not sufficient. Held-out validation thus indicates not only which current model to select, but also what kind of new information is needed for improvement.

\section{Discussion and outlook}
\label{sec:discussion}

The first empirical finding is the emergence of scale-stationary normalized cloud shapes in a non-smooth velocity field. This is surprising because the cloud remains subject to multiplicative deformation, yet neither the minimum-area enclosing ellipse nor the mass ellipse shows continuing aspect-ratio growth over the extended post-transient interval. Saturation is statistical and broad: individual clouds continue to fluctuate, while their scale-conditioned shape distributions become approximately stationary.

Using both ellipses is essential. The mass ellipse follows a more anisotropic particle-weighted core. The outer ellipse follows a larger region, is determined by boundary particles, and has a more stable conditional shape distribution. It is therefore not merely a noisier covariance ellipse; it describes a different geometric layer of the same cloud. This paired description connects naturally to principal-component diagnostics being developed for dense particle clouds in convection \cite{ettel_lagrangian_2026}.

The second empirical finding is the shape-production/correction balance. A gradient averaged over the enclosing region and an affine map fitted to the particle velocities are different coarse-grained objects. In the present experiment, their difference explains most of the negative correction to aligned stretching. The residual beyond the best affine particle map is smaller and specific to the outer-envelope representation. The relative sizes of these terms are not expected to be universal; the transferable result is the decomposition and the ability to measure each part separately.

The reduced-model exercise shows how the geometric measurements can be converted into a testable stochastic description. It complements recent data-driven models of Lagrangian velocity-gradient dynamics \cite{das_data-driven_2024,carbone_tailor-designed_2024} by augmenting the perceived gradient with finite-cloud geometry and by selecting model structure on held-out realizations. M2 is the smallest model in the tested hierarchy that reproduces both one-step statistics and mean shape evolution on held-out trajectories. The failure of the fully coupled M3 model to improve prediction is equally informative: more coefficients acting on the same instantaneous state are not enough. Consistent with recent memory-based Lagrangian reduction \cite{dewit_mori-zwanzig_2026}, future models should first test memory, additional geometric descriptors, or non-Gaussian conditional increments rather than simply enlarge the drift matrix.

This is what we mean by a physics-informed data methodology. Geometry determines the variables; exact kinematics determines which terms are kept; data determine residual coefficients and supported couplings; and independent trajectories determine whether added structure is justified. The same sequence can be applied to direct numerical simulations or experiments even though the selected equations will change. In three dimensions, each mass or enclosing ellipsoid has three principal semiaxes and therefore two independent logarithmic aspect ratios after overall scale is removed. The relative orientation of the ellipsoid and the five-component symmetric-traceless strain is an $SO(3)$ object rather than one doubled angle. Nevertheless, the central distinction among a region-averaged gradient, a particle-fitted affine map, and an outer-envelope correction remains well defined. Realistic flows may also require conditioning on local flow structures, pressure-related variables, intermittency, or explicit memory.

The synthetic scope is both a limitation and a deliberate design choice. The present velocity has prescribed Gaussian statistics and no Navier--Stokes dynamics. Existing tetrahedron studies in homogeneous turbulence and dense-cloud studies in convection show that multi-particle geometry is measurable and physically informative \cite{biferale_multiparticle_2005,ettel_lagrangian_2026}, but they do not settle the saturation result reported here. The next applications should treat saturation as an empirical question: whether normalized bulk and outer shapes become scale stationary, how the production/correction balance changes, and which state variables are needed for held-out prediction in each actual flow.

\section*{Acknowledgments}

This paper is dedicated to the memory of Misha Stepanov, who was tragically killed after being struck by a car on May 7, 2026. Over many discussions on Lagrangian closures, Misha emphasized the diagnostic value of evolving ellipsoidal summaries of particle clouds. His unpublished synthetic-flow notes \cite{stepanov_synthetic-flow_2024} provided an important motivation for the present work.

The author gratefully acknowledges financial support from the University of Arizona start-up program and prior support (2019--2024) from Los Alamos National Laboratory. The ideas developed here grew from the University of Arizona--Los Alamos ``MachinE Learning for Turbulence'' collaboration and from discussions with M. Stepanov, D. Livescu, C. Fryer, Y. Tian, M. Woodward, and C. Hyett.

Development of this work took place largely in May 2026 while the author was visiting Oak Ridge National Laboratory. The author gratefully acknowledges support through the laboratory's mini-sabbatical program and especially the hospitality and encouragement of J. Restrepo and R. Archibald.

Further discussions during a June-July 2026 visit to the Technische Universit\"at Ilmenau, supported by an Alexander von Humboldt Foundation fellowship, significantly sharpened the distinction between the mass and outer-envelope ellipses and the questions to be tested in realistic flows. The author thanks J. Schumacher, M. Ettel, and R. J. Samuel for these discussions and for exploring applications of the methodology to direct numerical simulations of turbulent Rayleigh--B\'enard convection.

Language and coding assistants, including Claude (Anthropic) and ChatGPT (OpenAI), were used for editorial and software-organization support. The author derived, checked, and takes responsibility for all mathematical arguments, code, data, numerical results, and conclusions.

\section*{Code and data availability}

The accompanying reproducibility package available at \url{https://github.com/mchertkov/LagrangianEllipsoid2} contains the production data, Python scripts, executed Jupyter notebooks, and machine-readable tables used for all figures and quoted numbers.


%

\end{document}